\documentclass[a4paper,12pt,english,preprint,nofootinbib]{revtex4-1}
\usepackage{setspace}
\usepackage{float}
\usepackage{graphicx}
\usepackage{rotating}
\usepackage{caption}
\usepackage{subfig}
\usepackage[countmax]{subfloat}
\usepackage{array}
\usepackage{amssymb,amsmath}
\usepackage{bm}
\usepackage{xcolor}

\begin{document}

\title{Analytical  calculation of  pressure for confined atomic and molecular systems using the eXtreme-Pressure Polarizable Continuum Model}
\author{Roberto Cammi,$^{*}$}
\affiliation{Department of Chemical Science  Department of Chemical Science, Life Science and Environmental Sustainability , 
University of Parma, I-43100 Parma, Italy}
\author{Bo Chen}
\affiliation{Department of Chemistry and Chemical Biology,Baker Laboratory,Cornell University, Ithaca NY 14853-1301 (USA)}  
\author{Martin Rahm}
\affiliation{Department of Chemistry and Chemical Engineering, Division of Chemistry and Biochemistry, Chalmers University of Technology, Gothenburg, Sweden} 

\date{v8.0}
\today

\begin{abstract}
We show that the pressure acting on atoms and molecular systems  within the compression cavity of the eXtreme-Pressure Polarizable Continuum method can be expressed in terms of the electron density of the systems and of the Pauli-repulsion confining potential. The analytical expression holds for spherical cavities as well as for cavities constructed from van der Waals spheres of the constituting atoms of the molecular systems.

\end{abstract}
\maketitle
\newpage
\section{Introduction}

This paper presents a new development of the  eXtreme-Pressure Polarizable Continuum Model (XP-PCM), a Quantum Chemical method recently developed for the description of molecular systems at high pressure in condensed phases \cite{Cammi2008,Cammi2012,Cammi2015,Chen2017} and  already applied to study the effects of pressure on the electron density and the equilibrium geometry \cite{Cammi2012,Pagliai2014,Pagliai2016,Caretelli2017,Cammi2018},  on the  IR/Raman vibrational frequencies \cite{Cammi2012,Pagliai2014,Pagliai2016,Caretelli2017,Cammi2018},  on the electronic excitation energy \cite{Fukuda2015}, and on chemical reaction energy profiles \cite{Cammi2015,Chen2017,Yang2017,Cammi2017}.

The XP-PCM method describes the effects of the pressure by confining a molecular system within a cavity of an external transmitting medium that it is characterized by a dielectric permittivity and an averaged electronic charge density, both depending on the given condition of pressure. 
Forced within the cavity, the molecule has tails of its electronic distribution that can penetrate the cavity walls overlapping with the electron distribution of the surrounding medium and giving rise to a Pauli repulsive and confining interaction.
Pressure enters through this Pauli repulsive interaction between the molecule and the external medium, with the pressure that increases with the increase of the Pauli repulsion. Operatively,   the increase of the pressure is modeled by a suitable shrinking of the molecular cavity and by a correlated increase of the electron density of the medium {\cite{XP-PCMlimits}}.

The new development of the XP-PCM we are presenting regards the analytical calculation of the pressure acting on the confined atomic/molecular system.
The pressure is defined as minus the derivative of the electronic energy of the molecule in the solvent field with respect to the volume  of the confining cavity.   So far this derivative has been evaluated only by numerical methods \cite{Cammi2012,Fukuda2015}.
An analytical calculation of the pressure  is attractive from a theoretical as well as a computational point of view.
From a theoretical perspective it satisfies our desire to discover new and unexpected relationships between physical observables, hence strengthening  a unified view of the phenomenon under examination. 
From a computational point of view an analytical approach offers a more stable and accurate approach to the calculation of physical observables, hence opening new possibilities of application for the underlying theoretical methodology. 

In particular, our analytical theory of the pressure has been  motivated by a study of the properties of compressed atoms of all the elements (1-96) of the periodic table,  whereas numerical methods  for the calculation of the pressure have failed  to account for   sudden changes  in the electronic configuration of compressed atoms. We have first developed  an analytical theory for the case of atomic systems confined by simple spherical cavities. Then we have generalized it, in the form we are  presenting now, to the case of  cavities constituted from superimposed spheres (i.e. van der Waals cavities) used  for the compression of molecular systems.

\section{The XP-PCM theory: electronic energy, pressure and cavity step function}
The electronic energy, $G_{er}$, for a molecular system confined within a XP-PCM cavity is given by:                                                                                      
\begin{equation}
 G_{er}(p,\mathbf R)=<\Psi|\hat H^o+\frac{1}{2}{\hat{V}}_e(\Psi)+\hat V_{r}|\Psi>,
 \label{Ger}
\end{equation}
where $|\Psi>$ is the electronic wave-function, $H^o$  the Hamiltonian operator of the isolated molecule,   $\hat V_{e}({\Psi})$ and  $\hat V_{r}$  operators representing, respectively, the  electrostatic  and the  Pauli repulsion interactions between the molecule and the external medium.  \cite{Tomasi2005,Amovilli1997,Ve}.
 
The Pauli repulsion operator $\hat V_{r}$ of Eq. (\ref{Ger}), a key physical component of the XP-PCM model, corresponds to  a  repulsive step potential located at the boundary of the cavity enclosing the molecular solute \cite{Amovilli1997}:
 \begin{equation}
 \hat V_{r}=\int {\hat \rho}(\mathbf{r}){\Gamma}(\mathbf{r})d\mathbf{r},
\label{Vr}
\end{equation} 
 where ${\hat \rho}(\mathbf{r})=\sum_i^N\delta (\mathbf{r}-\mathbf{r}_i)$ is  the electron density operator (over the $N$ electrons of the molecular system)  and $\Gamma(\mathbf{r})$ a step barrier potential at the boundary of the molecular cavity:
\begin{eqnarray}
 \Gamma(\mathbf{r})={\cal Z}\Theta_C(\mathbf{r}) 
 \qquad \Theta_C(\mathbf{r})= \left \{
\begin{array}{l l}
0 & \mathbf{r}\subseteq \mathbb{D}_C \\
1 & \mathbf{r}\nsubseteq \mathbb{D}_C \\
\end{array}
\right. .
\label{Gamma}
\end{eqnarray} 
In Eq. (3),   ${\cal Z}$ is the height of the potential barrier and $\Theta_C(\mathbf{r})$ the cavity step function,  a Heaviside unit step function having a value equal to zero inside  the cavity  and equal to one outside of it ($\mathbb D_C$ denotes the domain of the physical space inside the cavity). The height of the potential barrier $\cal Z$   depends on the mean valence electron density $\rho_S$ of the external medium, estimated  at the given condition of the pressure $p$. \cite{Cammi2012}. Here, we note that an alternative form of the Pauli repulsion operator $V_r$ has been proposed by  Chipman and co-workers \cite{VrChipman-1,VrChipman-2}.

{{
The Schr\"{o}dinger equation which determines the electronic  wave-function $|\Psi>$ of the compressed molecular  systems is given by 
	\begin{equation*}
	\left [\hat H^o+{\hat{V}}_e(\Psi)+\hat V_{r}\right ]|\Psi>=E|\Psi>
	\label{Schrodinger}
	\end{equation*} 
and it solutions can be obtained at the various levels of Quantum Chemisty (Hartree-Fock, DFT, Coupled-Cluster,..).}}

The electronic energy $G_{er}$ is central in the study of the properties of compressed atoms and molecules. In fact, the derivatives of the electronic energy $G_{er}$ with respect to suitable parameters describing  perturbing agents \cite{Cammi2018} determine the various electronic properties of compressed systems. Furthermore, $G_{er}$  acts as effective potential  for the nuclei and the minima of the corresponding potential energy surface (PES) $G_{er}(\textbf{R})$ correspond to equilibrium geometries of a compressed  molecules. For a fast and effective exploration of the PES analytical derivatives of $G_{er}$ with respect to the Cartesian coordinates of the atomic nuclei  has been proposed in Ref. \cite{Cammi2012} and systematically applied to the study of equilibrium geometries and vibrational frequencies of compressed molecules \cite{Cammi2012, Pagliai2014,Pagliai2016,Caretelli2017}.

\subsection{The pressure}
As mentioned in the Introduction, the pressure  is defined as the change of the electronic energy  $G_{er}$ under variation of  the volume, $V_c$, of the molecular cavity:   
\begin{equation}
\label{p}
p=-\left(\frac{\partial G_{er}}{\partial V_{c}}\right).
\end{equation}  
The volume  $V_c$ pertains to a molecular van der Waals cavity constructed from a set of spheres centered on the nuclei of the constituting atoms. The spheres have radii, $R_i=R_{i}^0f$, $R_{i}^0$ being the atomic van der Waals radii \cite{Pauling1960,Bondi1964,Batsanov2001,Mantina2009,Alvarez2013,Rahm2016}, and $f$ being a scale factor. 
An upper value of the scaling factor, $f_0$, is set as a reference  and lower values of  $f$ are used to decrease the volume $V_c$ and hence increase the pressure $p$.
The reference scaling factor $f_0$ depends on the chosen set of van der der Waals atomic radii.The value $f_0=1.2$ is commonly used \cite{Cammi2012} in combination with radii as tabulated by Bondi \cite{Bondi1964}, while $f_0=1.3$ is used in combination with a new set of radii for the elements (1-96) of the atomic table as proposed by one of the present authors \cite{Rahm2016}. {{ We note that in general the volume of the cavity scales as a cubic polynomial in $f$.}}

{{
The XP-PCM definition of the pressure given in Eq. (4)  is equivalent to  the definition of   pressure given for simple quantum confining  models  (\textquotedblleft box models\textquotedblright), that have been used since the fourth decade of the past century for the study of the effect of the very high pressure on the properties of atoms and simple diatomic molecules \cite{Michels1937,Sommerfeld1938,tenSeldam1952,Byers-Brown1955,Ludena1978,Ley-Koo1979, LeSar1981,LeSar1983,Jaskolski2000,Connerade2000,Banerjee2002,Sen2005a,Sen2005b,Aquino2006,Guerra2009,Garza2009,Katriel2012,Sen2014,Rodriguez2015}. In these models a atom or a simple diatomic is collocated in a spherical or ellipsoidal box having an impenetrable (or even penetrable) confining boundary potential. To model compression on the quantum system the size of the box is decreased  and for a give size the  pressure is computed as the minus the derivative of the electronic  energy  of the system with respect to the volume fo the box, as in our Eq. (4).

It is worth to recall other two approaches for the calculation of the pressure based on a quantum mechanical calculations on  atoms or molecules.
The first approach has been proposed by R. F. Bader, and it is rooted in the definition of atoms in molecules as open systems bounded by a surface of zero flux of the gradient vector field of the electron density \cite{Bader1990}. In this approach the pressure  acting on a given open system is defined as the quantum force exerted per unit area of the surface  enclosing the open system \cite{Bader1997,Bader2005}, then the pressure may be determined exploiting the virial theorem for open quantum systems or as the derivative of the energy of the open quantum systems with respect to its volume. When an entire molecule is considered as a quantum system this definition of pressure corresponds formally to Eq. (4) \cite{Bader2009}.

The second approach is based the Weisskopf's kinetic energy pressure concept of steric repulsion \cite{Weisskopf1975}, as expressed within the Natural Bond Orbital (NBO) analysis of Weinhold \cite{Weinhold1997a}. In this case, the pressure may be evaluated by encircling the target molecule with a cluster of noble gas and computing the NBO steric repulsion. The formal connection with the XP-PCM  pressure implies to consider the derivative of the steric repulsion energy with respect to a suitable definition of the volume of the target molecular system  \cite{Weinhold1997b}. This NBO approach for the calculation of the pressure may be of particular interest as an alternative way to calibrate the boundary confining barrier of the XP-PCM model.}}

\subsection{The van der Waals cavity step function: the cavity volume and the Pauli repulsion operator}
For a molecular van der Waals cavity  the cavity step function $\Theta_C $ of Eq. (3) assumes a simple analytical form.
Let us consider a van der Waals cavity constituted by the superposition of $N_S$ atomic van der Waals spheres with centers $\left\lbrace {\mathbf C}_i \right\rbrace $ and radii $\left\lbrace R_i=fR_i^0 \right\rbrace $ (see Fig. 1). The corresponding
cavity  step function can be expressed as 
\begin{eqnarray}
 \Theta_{C}({\mathbf r};f)=\prod_{i=1}^{N_S}\Theta_i({\mathbf r};{\mathbf C}_i,fR_i^0),
\end{eqnarray} 
where the product extends over the  $N_S$ atomic spheres and the $\Theta_i({\mathbf r};{\mathbf C}_i,fR_i^0)$ 
 are  spherical Heaviside step functions:
\begin{equation}
 \Theta_i\left( {\mathbf r};{\mathbf C}_i, fR_i^0\right)=\left \{
\begin{array}{l l}
0 & \mathbf{r}\subseteq \mathbb{D}_i \\
1 & \mathbf{r}\nsubseteq \mathbb{D}_i .\\
\end{array}
\right.
\end{equation}
where $\mathbb{D}_i $ are the domains of space internal to the spheres.

In terms of the van der Waals cavity step function (5) the volume of the cavity,$ V_{C}(f)$ is given by 
\begin{equation}
 V_{C}(f)=\int \left( 1-\prod_{i=1}^{N_S}\Theta_i({\mathbf r};{\mathbf C}_i,fR_i^0)\right) d \mathbf r ,
 \end{equation}
 
and  the Pauli repulsion operator $\hat V_{r}(\mathbf{r};f)$ can be rewritten as:
 \begin{equation}
 \hat V_{r}(\mathbf{r};f)={\cal Z}(f) \int {\hat \rho}(\mathbf{r})\prod_{i=1}^{N_S}\Theta_i({\mathbf r};{\mathbf C}_i,fR_i^0)d\mathbf{r} ,
\label{Vr_f}
\end{equation} 
where ${\cal Z}(f)$ is given by \cite{Cammi2015,Zf}
\begin{equation}
\label{Zf0}
{\cal Z}(f)= {\cal Z}_0\left (\frac{V_c(f)}{V_c(f_0)}\right)^\frac{-(3+\eta)}{3} .  
\end{equation}
In Eq. (9),  ${\cal Z}_0$ is the step barrier  at the standard condition of pressure \cite{Amovilli1997}, and $\eta$  a semi-empirical parameter that gauges how strong  the Pauli repulsive  barrier of the external medium. From Eq. (9) it follows that Pauli barrier increases  as $\eta$  increases and therefore a higher $\eta$ denotes a harder (and less compressible) external medium. 

The value of the Pauli repulsion parameter $\eta$ can be estimated by comparison of the computed pressure-Volume results of XP-PCM  with available experimental  pressure-Volume data \cite{Cammi2012,Pagliai2014,Fukuda2015}.  A value of $\eta=6$ gives a dependence of the computed  pressure (4) on the cavity volume $V_C$ in reasonable agreement with the dependence of the experimental pressure on the molar volume in molecular solids. {{We will give an example of such a numerical comparison in Section \textbf{IV} for the case of a compressed argon atom. We note, as historical remark, that the calibration of a boundary potential of a compressed quantum system by  comparison with the experimental pressure-Volume data  has been first suggested  by R. LeSar and  D. R. Herschbach in their seminal paper \cite{LeSar1981} on a confining spheroidal box model for the description of the effect of the pressure on the electronic and vibrational properties of the hydrogen molecule.}}


\section{Analytical theory of the pressure}
Using the cavity scale factor $f$ as the operative parameter to gauge the volume of the cavity, the expression of the pressure in  Eq. (4) should be rewritten using the chain rule of differentiation as :  
\begin{equation}
\label{pressure}
p=-\left(\frac{\partial G_{er}}{\partial f}\right)\left(\frac{\partial V_{c}}{\partial f}\right)^{-1},
\end{equation}
where $ \partial G_{er}/\partial f $ is the derivative of the electronic energy functional with respect to the cavity scaling factor $f$, and $ \partial V_{c}/\partial f $ the corresponding derivative of the volume of the cavity.


\subsection{The derivative of the volume of the cavity  with respect to the scaling factor f} 


A direct differentiation of Eq. (7) with respect to the cavity scaling factor $f$ gives 
\begin{equation}
  \frac{d V_{c}(f)}{df}=   -\int \sum_{i=1}^{N_S} \frac{d \Theta_i({\mathbf r};{\mathbf C}_i,fR_i^0)}{df} \prod_{j\neq i}^{N_S-1}\Theta_j({\mathbf r};{\mathbf C}_i,fR_i^0)d{\mathbf r} .
 \end{equation}
The derivative of the  i-th atomic spherical Heaviside function  of Eq. (11) is  given by
\begin{equation}
\frac{d \Theta_i({\mathbf r};fR_i^0)}{df} =-R_i^0\delta\left( |{\mathbf r}-{\mathbf C}_i|-fR_i^0\right) ,
\end{equation}
where $\delta$ is the unit impulse function (or Dirac delta function); hence the derivative  of the cavity volume with respect to the cavity scaling factor $f$ may be expressed as
\begin{equation}
 \frac{d V_{c}(f)}{df}= \sum_iR_i^0 S_i, 
 \end{equation}
where $S_i$ denotes the contribution of the i-th atomic sphere to the total cavity surface $S_C$ (see Fig. 1):
\begin{equation}
S_i=\int \delta\left( |{\mathbf r}-{\mathbf C}_i|-fR_i^0\right) \prod_{j\neq i}\Theta_j({\mathbf r};{\mathbf C}_i,fR_j^0)d{\mathbf r} .
\end{equation}
The surface $S_i$ can also be expressed in terms of  spherical coordinates  as: \cite{Polar}
\begin{equation}
 S_i= \int_0^\pi\int_0^{2\pi}  \left (fR_i^0\right )^2 \prod_{j\neq i}\Theta_j(fR_i^0,\theta,\phi;fR_j^0)sin \theta d{\theta}d\phi ,
\end{equation}
where the product of the spherical step functions is non-zero only if the point on the i-th sphere, with polar coordinates  $fR_i^0,\theta,\phi$, is outside of all other van der Waals spheres (i.e. if the point is on the boundary of the cavity and not on its interior)\cite{Pomelli2007}.

 \subsection{The  derivative of the electronic energy with respect to the cavity scaling factor f}

 This derivative of the electronic energy $G_{er}$ can be evaluated by applying an extension of the Hellmann-Feynman theorem to the XP-PCM model \cite{Cammi2013,Cammi2014}. According to this extension the derivative of the electronic energy $G_{er}$ with respect to the cavity scaling factor $f$ is given by \cite{Cammi1994} : 
\begin{equation}
\label{pressure}
\left(\frac{\partial G_{er}}{\partial f}\right)= <\Psi|\frac{\partial \hat V_r}{\partial f}|\Psi> ,
\end{equation}
where $\frac{\partial \hat V_r}{\partial f}$ is the partial derivative of the  Pauli repulsion operator $\hat V_r$ of Eq. (8) with respect to the cavity scaling factor $f$. 


A direct differentiation with respect to the cavity scaling factor $f$ of $\hat V_r$  leads to the following form of the derivative of the electronic energy $G_{er}$: 
\begin{equation}
\left(\frac{\partial G_{er}}{\partial f}\right)= \frac{d{\cal Z}(f)}{df}\int {\rho}(\mathbf r)\Theta_C({\mathbf r};f) d\mathbf r+{\cal Z}(f)\int {\rho}(\mathbf r) \frac{d }{df} \prod_{i}\Theta_i({\mathbf r};{\mathbf C}_i,fR_i^0)d\mathbf r .
\end{equation}

Here, the first term on the right side represents the contribution to the pressure due to the dependence of the Pauli repulsion barrier ${\cal Z}(f)$ on the cavity scaling factor (see Eq. 10), while the second term represents the contribution to the pressure due to the contraction of the boundary of the cavity determined by the cavity scaling factor $f$.

\subsection{The  analytical expression fo the pressure}

A direct differentiation of Eq. (9) gives the following expression for  the derivative of the Pauli repulsion barrier in the first term on the right side of Eq. (17): 
\begin{equation}
\frac{d{\cal Z}(f)}{df}=\frac{-(3+\eta)}{3V_c(f)}\frac{d V_c(f)}{df}{\cal Z}(f) ,
\end{equation}
where $\frac{d V_c(f)}{df}$ is the derivative of the cavity volume; hence  by introducing Eq. (15)  we obtain:
\begin{equation}
\frac{d{\cal Z}(f)}{df}=\frac{-(3+\eta)}{3V_c(f)}\left ( \sum_iR_i^0 S_i \right){\cal Z}(f) .
\end{equation}

The derivative of the cavity step function within the integral of the second term of the right side of Eq. (17)  can be rewritten as 
\begin{equation}
\int {\rho}(\mathbf r)\frac{d}{df}\prod_{i}\Theta_i({\mathbf r};R_i^0,f) d\mathbf r=\sum_i R_i^0S_i[\rho] ,
\end{equation}

where $S_i[\rho]$ is give by:
\begin{equation}
 S_i[\rho]= \int_0^\pi\int_0^{2\pi}  \rho(fR_i^0,\theta,\phi)\left (fR_i^0\right )^2 \prod_{j\neq i}\Theta_j(fR_i^0,\theta,\phi;fR_j^0)sin \theta d{\theta}d\phi .
\end{equation}


From Eq.s (18-21) it follows that the derivative of $G_{er}$ with respect to the cavity scaling factor can be expressed as:
\begin{equation}
\left(\frac{\partial G_{er}}{\partial f}\right)= \left (\frac{-(3+\eta)}{3V_c(f)} \right )\left ( \sum_iR_i^0 S_i \right){\cal Z}(f)\int {\rho}(\mathbf r) \Theta_{C}(\mathbf r;f)d\mathbf r+{\cal Z}(f)\sum_i R_i^0 S_i[\rho] .
\end{equation}

Finally, by substituting Eq.s (13) and (22) into Eq. (10) the pressure $p$ is given by
\begin{eqnarray}
p= \frac{(3+\eta)}{3V_c(f)} {\cal Z}(f)\int {\rho}(\mathbf r) \Theta_{C}(\mathbf r;f)d\mathbf r - {\cal Z}(f)\frac{\sum_i R_i^0S_i[\rho]}{\sum_i R_i^0S_i} .
\end{eqnarray}

This analytical expression relating the pressure on atomic/molecular systems within the XP-PCM method to their electron density ${\rho}(\mathbf r)$ and to the confining potential ${\cal Z}(f)$  represents the key theoretical development of this paper. 

We remark that in the case of a spherical cavity the analytical expression of pressure of Eq. (23) reduces to the simplified form 
\begin{equation}
 p = 
 \frac{(3+\eta)}{ 4\pi \left(fR_0\right )^3}{\cal Z}(f)\int {\rho}(\mathbf{r}){\Theta_{i}}(\mathbf{r})d\mathbf{r} -{\cal Z}(f)\frac{S_i[\rho]}{4\pi \left(fR_0\right )^2} .
\end{equation}


\section{Numerical results}
In this  section we present the results of applying our analytical theory for the calculation of the pressure in the XP-PCM method.

The analytical pressures  are compared  with those obtained from two pure numerical methods.
The first numerical method is based on a simple finite difference approximation of the derivatives of the electronic energy with respect to the volume of the cavity in Eq. (4) . We use a central difference method  and the pressure is evaluated as
\begin{equation}
 p\backsimeq -\frac{G_{er}(f+\Delta f)-G_{er}(f-\Delta f)}{V_{c}(f+\Delta f)-V_{c}(f-\Delta f)}, 
\end{equation}
where $\{f_i\}$ is a reference cavity scaling factor  and $\Delta f$ is a suitably small variation of it. 

The second numerical method is based on an analytical fitting  of the electronic energy $G_{er}(f)$ as a function of the cavity volume  $V_c(f)$. Electronic  energies and cavity volumes are computed  for a selected set of values  $\{f_i\}$ of the cavity scaling factor, and the fitting is performed by using  the  following functional form \cite{Fukuda2015}:
\begin{equation}
 G_{er}(V_v)\backsimeq G_{er}(V_c^0)+a*V_c\left[ \frac{1}{b-1}\left( \frac{V_c^0}{V_c}\right)^b+1\right]+c*V_c .
\end{equation}

Hence, a direct differentiation gives the pressure as a function of the volume of the cavity: 
\begin{equation}
 p=a*\left [ \left( \frac{V_c^0}{V_c}\right)^b-1\right]-c .
\end{equation}
The numerical fitting method encounters problems of numerical inaccuracy when the fitting functional form (26) is not adequate, as in the case of sudden variations of the electronic energy due to changes of the ground  state electronic configuration of the atomic or molecular system.  

\subsection{Computational details}
The XP-PCM calculation where carried out on an argon ($Ar$) and an acetylene molecule ($C_2H_2$). All the calculation have been performed at the DFT \cite{Kohn1965} level using the B3LYP exchange-correlation functional \cite{Becke1993} and the aug-cc-pVTZ basis set \cite{Kendal1992}. {{ For sake of simplicity, in the comparison of the three methods for the calculation of the pressure we have taken fixed all the others parameters of the calculation; in the case of the acetylene molecule we have therefore assumed for all the values of the pressure a fixed geometry, corresponding to the equilibrium geometry of acetylene in the gas phase \cite{geom}.}} Cyclohexane was used as the external transmitting medium ( $\rho_M=0.2004 e/\AA{}^3$, $\epsilon=2.0165$ at standard condition of temperature and pressure). A value of $\eta=6$ for the Pauli repulsion parameter  (cfr. Eq. (9)).  The  pertinent van der Waals cavities have been defined using the Bondi atomic  van der Waals radii \cite{Bondi1964} $(R_H^0=1.2\AA{},\quad R_C^0=1.7\AA{},\quad R_{Ar}^0=1.88 \AA{})$.
All the calculations were performed using a local version of the Gaussian 09 suite of programs  \cite{Gaussian}.

\subsection{An Argon atom compressed within a spherical cavity}

 The values of electronic energy $G_{er}$ of argon as a function of the cavity scaling factor $f$ and of the cavity volume $V_c$ are reported in Table I \cite{TSare0075}.  As expected, the electronic energy increases as the volume of the cavity decreases; a nonlinear fitting with  the analytical form (26) of the electronic energy $G_{er}$ as a function of the cavity volume $V_c$ gives the following values of the fitting  parameters: $ a=(3.25\pm0.11)*10^{-4} Hartree/\AA{}^3,\quad b (4.139\pm0.044),\quad c=-(4.36\pm0.29)*10^{-4} Hartree/\AA{}^3$.
\begin{table}[H]
  \caption{Electronic energy $G_{er}$ (Hartree) for an argon atom in a XP-PCM spherical cavity as a function of the cavity scaling factor $f$ and cavity volume $V_C$ ($\AA{}^3$). }
  \centering
\begin{tabular}{|c|c|c|}
\hline \hline
 \emph{f} & $V_c$ &	 $G_{er}$   \\
\hline
1.2	&	48.096	&	-527.559602330	\\
1.15	&	42.331	&	-527.558750538	\\
1.1	&	37.046	&	-527.557027784	\\
1.05	&	32.220	&	-527.553522729	\\
1	&	27.833	&	-527.546360053	\\
0.95	&	23.863	&	-527.531687868	\\
0.925	&	22.029	&	-527.519335543	\\
0.9	&	20.290	&	-527.501677093	\\
0.875	&	18.646	&	-527.476504211	\\
0.85	&	17.093	&	-527.440710191	\\
\hline \hline
\end{tabular}
\end{table}

The values of the pressure $p$ as function of cavity volume $V_c$ are reported in Table II for the three different methods: the finite difference method (24) , the analytical method (23) and the numerical fitting method (27).  The values of the finite difference method have been obtained using a finite step of the cavity scale factor $f=\pm 0.005$  and the values of the fitting method have been obtained using the pertinent fitting parameters $a,b,c$.
\begin{minipage}{12cm}
\begin{table}[H]
  \caption{Pressure $p$ (GPa) as a function of the cavity scaling factor $f$ and volume $V_c$ ($\AA{}^3$)  for an argon atom in a XP-PCM spherical cavity, as computed using (Eq. 23), the finite difference  method (Eq. 25) and the fitting method (Eq. 27).}
  \centering
\begin{tabular}{|c|c|c|c|c|}
\hline \hline
 &  &	\multicolumn{3}{c|}{$p$}  \\
\hline
        &       &   \emph{Analytical}     & \multicolumn{2}{c|}{$Numerical$}	 \\ 
\hline
   \emph{f} & $V_c$ &   \emph{Eq. 23}     & \emph{Eq. 25}    & \emph{Eq. 27}	 \\ 
\hline
1.2	&	48.096	&	0.42	&	0.45	&	0.786	\\
1.15	&	42.331	&	0.94	&	0.94	&	1.339	\\
1.1	&	37.046	&	2.08	&	2.08	&	2.481	\\
1.05	&	32.220	&	4.65	&	4.65	&	4.926	\\
1	&	27.833	&	10.49	&	10.48	&	10.350	\\
0.95	&	23.863	&	23.81	&	23.81	&	22.875	\\
0.925	&	22.029	&	35.91	&	35.84	&	34.633	\\
0.9	&	20.290	&	54.15	&	54.57	&	53.112	\\
0.875	&	18.646	&	81.58	&	81.60	&	82.475	\\
0.85	&	17.093	&	122.68	&	122.74	&	129.794	\\

\hline \hline
\end{tabular}
\end{table}
\end{minipage}
\newline
Comparing  the finite difference and the analytical methods calculations, the results of Table II show very good agreement between these two methods. The differences between the analytical values and the numerical values are physically negligible, with a mean unsigned error between analytical values and the  finite difference values  of $\Delta p=0.06 \quad GPa$. 

Comparing the finite difference and the fitting methods calculations, the results of Table II show some significant differences, with a  mean unsigned error with respect to the finite difference values of $\Delta p=1.31 \quad GPa$.

Hence, for the case of spherical cavities the finite difference and the analytical methods give equivalent results, more accurate that those obtained from the fitting method. However, the analytical method is by far the most computationally effective: for a given value of the cavity scaling factor $f$, the pressure is computed by a single point QM calculation, while in the finite difference method the pressure requires additional QM calculations in nearby values of the cavity scaling factor $f$, requiring also care in the selection of the finite scaling factor step $\Delta f$. 

{{We close this section on the compressed argon atom with a comparison between the XP-PCM computed values of the  pressure as function of the volume compression with the corresponding experimental pressure-volume data of solid argon \cite{Exp-pV}. Table III shows the values of the pressure are expressed as a function  the compression  $V/V_1$; for the XP-PCM model $V$ is the volume of the cavity corresponding to the pressure $p$ and $V_1$ is a reference volume of the cavity corresponding to the pressure $p=1.1GPa$, while for the experimental data $V$ and $V_1$ are, respectively, the molar volume at the pressure $p$ and at the reference pressure $p=1.1GPa$. This comparison shows that the values of XP-PCM method are in good agreement with  the experimental data of pressure as a function  the volume compression.

\begin{minipage}{12cm}
	\begin{table}[H]
		\caption{{{Comparison of the pressure $p$ (GPa) as a function of the volume compression $V/V_1$ for a compressed argon atoms within the XP-PCM model and for the experimental solid argon  \cite{Exp-pV}. }}}
		\centering
		\begin{tabular}{|c|c|c|}
			\hline \hline
			 &	\multicolumn{2}{c|}{$V/V_1$ }  \\
			\hline
			  $p$   &   \emph{XP-PCM}     & \emph{Exp.}	 \\ 
			\hline
1	&	1	&	1	\\
2	&	0.900	&	0.918	\\
5	&	0.759	&	0.798	\\
10	&	0.678	&	0.705	\\
24	&	0.573	&	0.583	\\
36	&	0.529	&	0.528	\\
54	&	0.489	&	0.476	\\
82	&	0.448	&	0.426	\\
							\hline \hline
		\end{tabular}
	\end{table}
\end{minipage}
}}
\newline

\subsection{An Acetylene molecule compressed within a van der Waals cavity}

The values of electronic energy $G_{er}$ as a function of the cavity scaling factor $f$ and of the cavity volume $V_c$ for molecular acetylene are reported in table IV \cite{TSare0050}.  As expected, the electronic energy increases as the volume of the cavity decreases; the  nonlinear fitting (26) of the electronic energy $G_{er}$ as a function of the cavity volume $V_c$ gives the following values of the parameters: $ a=(2.3953 \pm 0.065)*10^{-4} Hartree/\AA{}^3,\quad b=(4.720 \pm 0.0495),\quad c=-(2.9845 \pm 0.128)*10^{-4} Hartree/\AA{}^3$.

\begin{table}[H]
  \caption{Electronic energy $G_{er}$ (Hartree) for an acetylene molecule in a XP-PCM van der Waals cavity as a function of the cavity scaling factor $f$ and cavity volume $V_c$ ($\AA{}^3$).}
  \centering
\begin{tabular}{|c|c|c|}
\hline \hline
 \emph{f} & $V_c$ &	 $G_{er}$ (Hartree)   \\
\hline
1.200	&	54.33	&	-77.365143 	\\
1.150	&	48.832	&	-77.363467 	\\
1.100	&	43.724	&	-77.360207 	\\
1.050	&	38.982	&	-77.354720 	\\
1.000	&	34.58	&	-77.344027	\\
0.950	&	30.489	&	-77.325121 	\\
0.925	&	28.566	&	-77.309103 	\\
0.900	&	26.701	&	-77.287310	\\
\hline \hline
\end{tabular}
\end{table}
The pressure $p$ for the acetylene molecule  as function of  cavity volume $V_c$ is shown in Table V. As for the case of the argon atom, the pressure has been computed with three different methods: the finite difference method (24), using with a finite step of the cavity scaling factor $f=\pm 0.002$, the analytical method (23) and the numerical fitting method (27). 
\newline

\begin{table}[H]
  \caption{Pressure $p$ (in $GPa$) as a function of the cavity volume $V_c$ (in $\AA{}^3$) and  of the cavity scale factor $f$ for molecular acetylene in a van der Waals cavity of cyclohexane using  (Eq. 23), the finite difference  method (Eq. 25) and the fitting method (Eq. 27).  }
  \centering
\begin{tabular}{|c|c|c|c|c|}
\hline \hline
 &  &	\multicolumn{3}{c|}{$p$ (GPa)}  \\
\hline
        &       &   \emph{Analytical}     & \multicolumn{2}{c|}{$Numerical$}	 \\ 
\hline
   \emph{f} & $V_c$ &   \emph{Eq. 23}     & \emph{Eq. 25}    & \emph{Eq. 27}	 \\ 
\hline
1.200	&	35.561	&	1.1     &	0.9	&	1.3	\\
1.150	&	31.299	&	2.2	&	1.7	&	2.2	\\
1.100	&	27.391	&	4.1	&	3.9	&	3.9	\\
1.050	&	23.823	&	7.8	&	9.3	&	7.2	\\
1.000	&	20.58	&	15.3	&	14.7	&	14.1	\\
0.950	&	17.644	&	29.7	&	30.5	&	28.7	\\
0.925	&	16.288	&	42.2	&	44.1	&	41.5	\\
0.900	&	15.002	&	60.0	&	58.8	&	61.0	\\
\hline \hline
\end{tabular}
\end{table}
The results of Table V show good agreement between  the finite difference and the analytical methods for the calculation of the pressure; the differences between the analytical and the numerical values correspond to  a  mean unsigned difference of $\Delta p=0.8 \quad GPa$.  Also the numerical fitting method gives similar results, with a mean unsigned differences with respect to the finite difference method of $\Delta p=1.1 \quad GPa$.

Hence, even for the case of a non spherical van der Waals cavity the numerical methods and the analytical methods are in good agreement. However, the analytical method is by far the most computationally effective for the same reasons discussed for a spherical confining cavity. 
\section{Conclusion and Perspectives}
We have presented an analytical theory for the calculation of the pressure for molecular systems confined in cavities with the XP-PCM method. The analytical theory  exceeds in efficiency and accuracy compared to pure numerical methodologies and increases the 
computational robustness of the XP-PCM methodology.
  
The analytical theory for pressure can be extended along several directions. One of them regards the quantum mechanical (QM) level for the description of the compressed molecular system . In this work we have assumed variational QM methods as Hartee-Fock or Density Functional Theory; however, the analytical theory can be easily generalized to non-variational methods like MP2 or coupled-cluster methods by exploiting the analytical derivative theories of the energy developed for these methods with the PCM solvation model \cite{Cammi1999,Cammi2009}. The theory can also be extended from a the time-independent to a time-dependent  Quantum Mechanical  description (i.e Real-time description) with the perspective of real-time computational study of molecular processes at extreme-high pressure \cite{Pipolo2016}.  

Another interesting generalization of the present theory regards the use of more complex confining cavities. If a van der Waals cavity presents interstitial spaces between the atomic spheres that are not accessible to the solvent molecules  the van der Waals surface does't accurately represent the boundary of the Pauli confining potential exerted by the external medium. In this case a more complex cavity based on the Solvent Excluding Surface (SES) \cite{Richards1977} is more physically  appropriate to represent the boundary of the Pauli confining potential. The extension of our theory to this type of cavities will be presented in a future paper.

\section{Acknowledgments}
The authors thank Prof. Roald Hoffmann for useful discussions and comments on the manuscript.

\clearpage
\section{Figures}
\begin{figure}[H]
 \centering
 \includegraphics[scale=0.55]{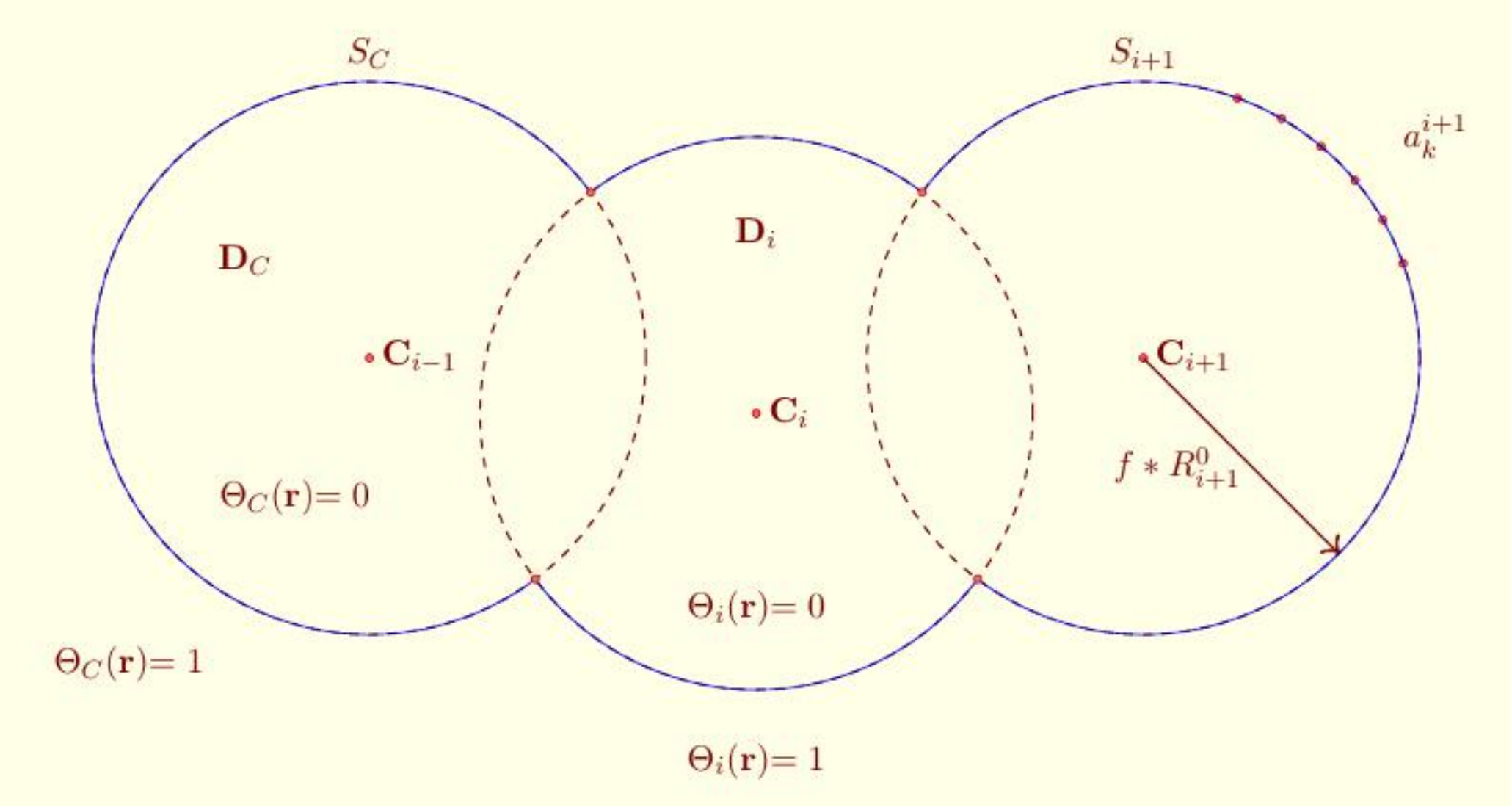}
 \caption{Schematic representation of a molecular van der Waals cavity. The atomic spheres have centers $\left\lbrace\mathbf C_i\right\rbrace$ and radii $\left\lbrace f*R_i^0 \right\rbrace$, being $f$ the cavity scaling factor and $R_i^0$ the reference atomic van der Waals radii (cfr. Eq. (5)). $\mathbf D_i$ denotes the domains of the atomic spheres and $\Theta_i$ the corresponding unit spherical step functions (cfr. Eq. (6)). $\mathbf D_{C}=\bigcup\left\lbrace\mathbf D_i\right\rbrace$ denotes the internal domain of the van der Waals cavity and $\Theta_{C}=\prod_i\Theta_i$ the corresponding unit step function (cfr. Eq. (3)) . $S_{C}=\sum_i S_i$ represents the area of the cavity surface, sum of atomic spheres contributions $S_i$ corresponding to the portions of the atomic spheres exposed to the solvent. Finally, the $\left\lbrace a_k^{i+1}\right\rbrace$ denote the area of the tesserae in which the atomic surfaces are partitioned, with $S_{i+1}=\sum_k a_k^{i+1}$.}  
 \label{fig:2}
\end{figure}


\begin{thebibliography}{99}
\bibitem{XP-PCMlimits}	{{In general, a  XP-PCM study requires an initial definition of the portion of matter to consider as the Quantum Mechanical active part and the definition of the chemically inert one. Furthermore an initial high end of the pressure range  has to be chosen. These operative aspects of the XP-PCM model must be decided on the basis of the nature of specific chemical problem under examination (spectroscopic problem, chemical reactions problem, etc ...).
.   
If necessary, the XP-PCM model has the capability to overcome the  limit of a chemically inert environment  by  explicitly including some selected molecular components of the environment as part of the active Quantum Mechanical portion.}}
\bibitem{Ve}{The argument of the electrostatic interaction operator ${\hat  V_{e}}(\Psi)$ denotes its non-linear character resulting from its functional dependence  on the electronic wave-function $\Psi$ of the molecular solute.}
\bibitem{Polar}{The polar spherical coordinates have center on the center ${\mathbf C}_i$ of the sphere.}:
\bibitem{Sri}{From the computational point of view, the exposed sphere surface  $S_i$ can be expressed in terms of the area of the tesserae in which the cavity surface is partitioned:
\begin{equation*}
\sum_i R_i^0\int_{{\mathbf r}_i\in S_i} {\rho}(\mathbf r_i)sin \theta d{\theta}d\phi=\sum_i R_i^0  \left ( \sum_k {\rho}(\mathbf r_i)a_k(i) \right) 
\end{equation*}}
\bibitem{TSare0075} All the results have been obtained using a tessellation of the surface of the cavity with an average area of tesserae $a_i=0.075\AA{}^2$.
\bibitem{TSare0050} All the results have been obtained using a tessellation of the surface of the cavity with an average area of tesserae $a_i=0.050\AA{}^2$.
\end{thebibliography}
\end{document}